\documentclass[letterpaper, 10 pt, conference]{ieeeconf}
\IEEEoverridecommandlockouts
\title{\LARGE \bf
Data-Driven Control with Inherent Lyapunov Stability
}
\author{Youngjae Min$^{1}$, Spencer M. Richards$^{2}$, Navid Azizan$^{1}$
\thanks{$^{1}$Youngjae Min and Navid Azizan are with the Laboratory for Information and Decision Systems, Massachusetts Institute of Technology, Cambridge, MA 02139,
USA.
        {\tt\small \{yjm, azizan\}@mit.edu}}%
\thanks{$^{2}$Spencer M. Richards is with the Autonomous Systems Laboratory (ASL), Stanford University, Stanford, CA 94305, USA
        {\tt\small spenrich@stanford.edu}}%
}

\let\proof\relax

\usepackage{amsthm}
\usepackage{amsmath}
\usepackage{amssymb}
\usepackage{mathtools}

\usepackage{microtype}
\usepackage{graphicx}
\usepackage{subfigure}
\usepackage{booktabs}

\usepackage{cite}
\newcommand{\citep}[1]{\cite{#1}}
\newcommand{\citet}[1]{\cite{#1}}

\DeclareMathOperator*{\argmin}{arg\,min}

\usepackage[bookmarks=true,citecolor=blue]{hyperref} 
\hypersetup{colorlinks=true,linkcolor=blue}

\usepackage[capitalize,noabbrev]{cleveref}

\theoremstyle{plain}
\newtheorem{theorem}{Theorem}[section]
\newtheorem{proposition}[theorem]{Proposition}

\theoremstyle{definition}
\newtheorem{definition}[theorem]{Definition}

\theoremstyle{remark}

\newcommand{\coils}{{\sf CoILS}}

\begin{document}

\maketitle
\thispagestyle{empty}
\pagestyle{empty}

\begin{abstract}
Recent advances in learning-based control leverage deep function approximators, such as neural networks, to model the evolution of controlled dynamical systems over time. However, the problem of learning a dynamics model and a stabilizing controller persists, since the synthesis of a stabilizing feedback law for known nonlinear systems is a difficult task, let alone for complex parametric representations that must be fit to data. To this end, we propose \emph{Control with Inherent Lyapunov Stability} (\coils{}), a method for jointly learning parametric representations of a nonlinear dynamics model and a stabilizing controller from data. To do this, our approach simultaneously learns a parametric Lyapunov function which intrinsically constrains the dynamics model to be stabilizable by the learned controller. In addition to the stabilizability of the learned dynamics guaranteed by our novel construction, we show that the learned controller stabilizes the true dynamics under certain assumptions on the fidelity of the learned dynamics. Finally, we demonstrate the efficacy of \coils{} on a variety of simulated nonlinear dynamical systems.
\end{abstract}

\section{Introduction}\label{sec:intro}
Data-driven approaches have shown notable successes in solving complex nonlinear control problems in robotics and autonomy, such as autonomous navigation, multi-agent control, and object grasping and manipulation \citep{hewing2020learning, liu2022motor, levine2018learning}. In learning-based control problems, the task of controller synthesis is often compounded with a lack of knowledge, or uncertainty, about the system dynamics. To this end, neural networks have been widely used for system identification from data prior to controller synthesis \citep{chen1990non, wang2017new, GuptaMendaEtAl2020, RichardsAzizanEtAl2021, RichardsAzizanEtAl2023}. However, when using such complex parametric representations to model dynamics, it can be challenging to provide any guarantees about the behavior of the learned system, particularly regarding the stability of the system under closed-loop feedback.

A traditional method for stabilizing nonlinear dynamical systems is linearizing the system dynamics around an equilibrium point and using the linear quadratic regulator (LQR) techniques to minimize deviation from that equilibrium. LQR methods can achieve closed-loop stability within a small region where the linear dynamics approximation is accurate, yet away from this region, they can fail spectacularly, particularly for highly nonlinear systems performing agile maneuvers \citep{SunRomeroEtAl2022}. Nonlinear controllers that are \emph{certified} to be globally stabilizing can be synthesized if a control Lyapunov function (CLF) for the system is known \citep{SlotineLi1991}. However, constructing a CLF even for known dynamics can be difficult to do exactly, and thus approximate approaches are popular. For example, polynomial approximations of the dynamics enable a search for sum-of-squares (SOS) polynomials as Lyapunov functions via semidefinite programming (SDP) \citep{TedrakeManchesterEtAl2010}. However, polynomial approximation can be a significant restriction on the class of function approximators used.

\subsection{Related Work} To this end, there has been substantial growth in literature on \emph{data-driven} learning of stability certificates for dynamical systems, particularly those modeled with complex parametric representations. One of the earliest of such works fits a Lyapunov function for a known uncontrolled dynamical system by penalizing violations of the corresponding Lyapunov decrease condition \citep{RichardsBerkenkampEtAl2018}. Learning certificate functions such as Lyapunov, barrier, and contraction metric functions for dynamical systems with \emph{sampled} point penalties or constraints on stability violations is a ubiquitous theme in the literature \citep{SindhwaniTuEtAl2018,ElKhadirVarleyEtAl2019,BoffiTuEtAl2020,KhosraviSmith2021,DawsonGaoEtAl2022}. However, when coupled with regression on unknown dynamics, such approaches do not even guarantee the learned model is stable. To resolve this issue, \citet{ManekKolter2019} recently proposed a method to jointly learn a stable uncontrolled dynamics model with a Lyapunov function. It guarantees stability of the learned dynamics model by restricting it to the stable halfspace described by the Lyapunov decrease condition. However, its naive extension to \emph{controlled} dynamics would restrict the learned model to be stabilizable by any control input, thereby hindering its ability to model general controlled nonlinear systems.

For controlled dynamics, the paradigm of sampled pointwise constraints largely persists in the literature. \citet{SinghRichardsEtAl2020} jointly learn a dynamics model and certificate to regularize the dynamics model to perform well over long time horizons with sampled linear matrix inequality (LMI) constraints, yet they do not learn any specific controller. \citet{SunJhaEtAl2020} assume the dynamics are known, and jointly learn a controller and a contraction metric certifying the stability of the closed-loop system with loss terms corresponding to sampled point violations of a stability inequality. For Lyapunov-based approaches, \citet{chang2019neural} jointly learn a controller and a Lyapunov function for known dynamics, while they actively add training samples that violate the Lyapunov decrease condition using a falsifier. \citet{zhou2022neural} extend this method to unknown control systems with some stability guarantees, but their method requires certain knowledge on the system such as its Lipschitz constant and linearized model around the origin. Moreover, their learned dynamics do not guarantee the existence of the Lyapunov function. \citet{KashimaYoshiuchiEtAl2022} directly apply the approach from \citet{ManekKolter2019} to the controlled case, albeit only for control-affine systems when the actuator matrix is known. All in all, efficient simultaneous learning of Lyapunov functions and nonlinear controllers for systems with completely unknown dynamics has remained an open problem in control and robotics.


\subsection{Contributions} In this work, we tackle the difficult task of learning stabilizing feedback controllers with proven guarantees for unknown nonlinear dynamical systems. To this end, we propose \emph{Control with Inherent Lyapunov Stability} (\coils{}), a new method for jointly learning a controlled dynamical systems model and a feedback controller from data, such that the model is \emph{guaranteed by construction} to be stabilized in closed-loop with the learned controller. \coils{} does this by simultaneously learning a parametric Lyapunov function, which is used to constrain the open-loop dynamics onto the subspace of dynamics stabilizable in closed-loop by the learned controller. We further show that, under certain assumptions on the fidelity of our learned dynamics model, the learned controller is also guaranteed to stabilize the true dynamics. Finally, we demonstrate the performance of our joint learning method in a number of controlled nonlinear dynamical systems.


\section{Problem Statement}
In this paper, we are interested in controlling the unknown nonlinear dynamical system
\begin{equation}\label{eq:system}
    \dot{x}(t) = f(x(t), u(t))
\end{equation}
with state $x(t)\in\mathcal{X}\subset\mathbb{R}^n$ and control input $u(t)\in\mathcal{U}\subset\mathbb{R}^m$ at time $t\in\mathbb{R}$. While we do not know the dynamics $f : \mathcal{X} \times \mathcal{U} \to \mathbb{R}^n$, we assume we have access to a finitely-sized dataset $\mathcal{D} = \{(x_i, u_i, \dot{x_i})\}_{i=1}^N$ of input-output measurements of the system~\eqref{eq:system}.

We want to \emph{jointly} learn the dynamics~$f$ and how to control them. Specifically, we want to \emph{stabilize} the system around an equilibrium point. The point $x_e\in\mathcal{X}$ is an \textit{equilibrium} of the closed-loop system $f_{u^*}(x) \coloneqq f(x, u^*(x))$ with feedback controller $u^*:\mathcal{X}\to\mathcal{U}$ if
\begin{equation}\label{eq:equilibrium}
    f_{u^*}(x_e) = f(x_e, u^*(x_e)) = 0.
\end{equation}
There are many types of stability; we summarize the pertinent ones in the definition below for uncontrolled and closed-loop systems, i.e., where the dynamics are a function of the state $x$ only.
\begin{definition}\label{def:stability}
    The system $\dot{x}=f(x)$ for $x\in\mathcal{X}$ is \textit{stable} at its equilibrium point $x_e\in\mathcal{X}$ if for any $\epsilon>0$, there exists $\delta_\epsilon>0$ such that $\|x(0)-x_e\|_2<\delta_\epsilon$ implies ${\|x(t)-x_e\|_2<\epsilon}$ for all $t\geq 0$. The system is \textit{asymptotically stable} at $x_e$ w.r.t.\ $B\subset\mathcal{X}$ if it is stable at $x_e$ and $\lim_{t\rightarrow\infty} \|x(t)-x_e\|_2 = 0$ for all $x(0)\in B$. The system is \textit{exponentially stable} at $x_e$ w.r.t.\ $B\subset\mathcal{X}$ if there exist $m, \alpha>0$ such that $\|x(t)-x_e\|_2\leq m\|x(0)-x_e\|_2e^{-\alpha t}$ for all $x(0)\in B$.
\end{definition}

For the controlled system~\eqref{eq:system}, we assume a feedback controller $u^* : \mathcal{X} \to \mathcal{U}$ exists such that the resulting closed-loop system $\dot{x} =  f_{u^*}(x)$ is exponentially stable at $x_e$ w.r.t.\ $\mathcal{X}$. Without loss of generality, we assume $x_e = 0$. Thus, our overall goal is to \emph{jointly} learn the dynamics~$f$ and an exponentially stabilizing feedback controller~$u^*$. As we discuss in \cref{sec:learning}, our approach relies on encoding this stabilizability by~$u^*$ in~$f$ by construction. However, for this purpose, the stability definitions in \cref{def:stability} are cumbersome to work with directly, so, we appeal to Lyapunov stability theory to introduce scalar-value functions that summarize stability properties of dynamical systems~\citep{sastry2013nonlinear, blanchini2008set}.
\begin{proposition}\label{prop:asympt}
    Consider the system $\dot{x}=f(x)$ where $x\in\mathcal{X}$. Suppose there exists a continuously differentiable function $V : \mathcal{X}\rightarrow\mathbb{R}$ that is positive definite (i.e., $V(x) > 0$ for $x \neq 0$ and $V(0) = 0$), and satisfies
    \begin{equation} \label{eq:lyapunov_cond_asympt}
        \nabla_{f} V(x) \coloneqq \nabla V(x)^\top f(x) < 0,
    \end{equation}
    for all $x\in\mathcal{X}\setminus\{0\}$. Then the system is asymptotically stable at $x=0$ w.r.t.\ $\mathcal{X}$.
\end{proposition}
Such a function $V$ is termed a \emph{Lyapunov function}. The key idea is that by condition~\eqref{eq:lyapunov_cond_asympt}, $V$ is decreasing along any trajectories generated by~$f$ and eventually converges to 0, which implies $x=0$. The existence of a Lyapunov function with additional properties is also a necessary and sufficient condition for exponential stability as follows.
\begin{proposition}\label{prop:exp}
    Consider the system $\dot{x}=f(x)$ where $x\in\mathcal{X}$. This system is exponentially stable at $x=0$ w.r.t.\ $\mathcal{X}$ if and only if there exists a continuously differentiable function $V : \mathcal{X} \to \mathbb{R}$ such that
    \begin{equation}
        c_1 \|x\|_2^2 \leq V(x) \leq c_2 \|x\|_2^2,\quad
        \nabla_{f} V(x) \leq -\alpha V(x),
    \end{equation}
    for all $x\in\mathcal{X}\setminus\{0\}$ and some constants $\alpha, c_1, c_2>0$
\end{proposition}
When \cref{prop:asympt,prop:exp} are restated for a closed-loop system $\dot{x}=f_u(x)$ with controller $u : \mathcal{X} \to \mathcal{U}$, the accompanying Lyapunov function~$V$ is also known as a \emph{control Lyapunov function (CLF)}. In the next section, we use \cref{prop:exp} to constrain a model of~$f$ to be exponentially stabilizable by a parametric controller $u^*$ as guaranteed by an accompanying parametric Lyapunov function~$V$.


\begin{figure}
    \centering
    \includegraphics[width=.45\textwidth]{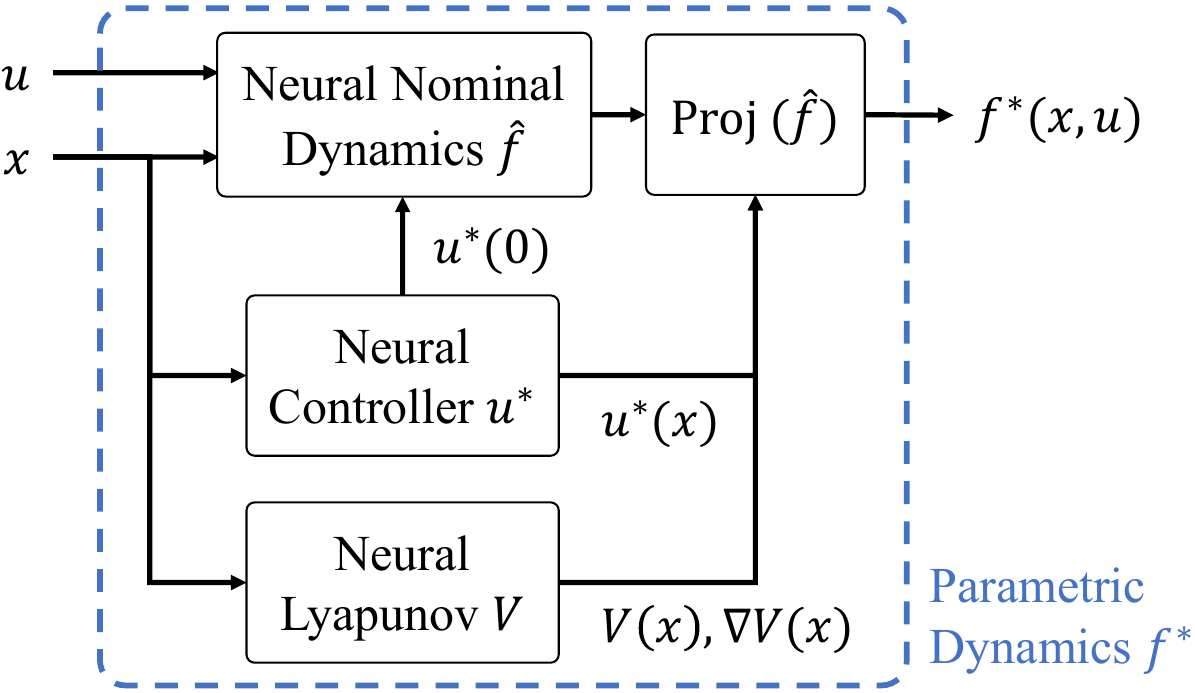}
    \caption{Schematic diagram of \coils{}. It projects the nominal dynamics model $\hat{f}$ to satisfy the stability by construction through incorporating the feedback controller $u^*$ and the Lyapunov function $V$ in the architecture.}
    \label{fig:model}
\end{figure}

\section{Joint Learning of Dynamics, Controller and Lyapunov Function}\label{sec:learning}
In this paper, we propose a novel architecture shown in Figure~\ref{fig:model} that satisfies the Lyapunov stability conditions by construction. Given the difficulty, if not infeasibility, of synthesizing a Lyapunov function for a separately learned dynamics model, we instead consider jointly learning them. Specifically, we propose a parameterization of the dynamics model that incorporates a given parametric Lyapunov function and controller. This connection allows us to jointly optimize them by applying automatic differentiation from a single loss function to fit the dataset.

More specifically, we construct the dynamics model $f^*$ by projecting a parametric \emph{nominal} model $\hat{f}:\mathcal{X}\times\mathcal{U}\rightarrow\mathbb{R}^n$ according to a parametric Lyapunov function $V : \mathcal{X} \rightarrow \mathbb{R}$ and a parametric feedback controller $u^*:\mathcal{X}\rightarrow\mathcal{U}$.
This projection constrains the dynamics model $f^*$ to satisfy the Lyapunov stability condition
\begin{equation}\label{eq:lyapunov_cond_exp}
    \nabla_{f^*_{u^*}} V(x)
    = \nabla V(x)^\top f^*(x,u^*(x)) \leq -\alpha V(x)
\end{equation}
for all $x\in\mathcal{X}\setminus\{0\}$ with a given $\alpha>0$. The construction of the dynamics model is as follows.
\begin{equation} \label{eq:fstar}
    f^*(x,u)
    \!=\! \begin{cases} \hat{f}(x,u) & \text{if } x \!=\! 0 \\
    \mathsf{Proj}_{u^*(x), V(x), \nabla V(x)}\left(\hat{f}(x,\cdot)\right)(u) & \text{o.w.}\end{cases}
\end{equation}
where
\begin{align}\label{eq:projection}
    &\mathsf{Proj}_{u^*(x), V(x), \nabla V(x)}\left(\hat{f}(x,\cdot)\right) \nonumber\\[1ex]
    &\begin{aligned}
    := &\hat{f}(x,\cdot) + \argmin_{\Delta f\in\mathbb{R}^n} & \hspace{-.3cm}\|\Delta f\|_2\hspace{4.6cm} \\[-1ex]
    & \hspace{2.05cm}\text{ s.t. } & 
     f^*(x,\cdot) \!=\! \hat{f}(x,\cdot) \!+\! \Delta f\hspace{2.3cm}\\
    & & \nabla V(x)^\top f^*(x,u^*(x)) \leq -\alpha V(x)\hspace{.65cm}
    \end{aligned}\nonumber\\[1ex]
    &= \hat{f}(x,\cdot) \!-\! \nabla V(x)\frac{\mathsf{ReLU}\bigl(\nabla_{}V(x)^\top\hat{f}(x, u^*(x)) \!+\! \alpha V(x)\bigr)}{\|\nabla V(x)\|_2^2}
\end{align}
provided $\nabla V(x)\neq 0$ for $x\neq 0$.

The interpretation of the projection is as follows. For all $x\in\mathcal{X}\setminus\{0\}$, if the nominal model $\hat{f}$ satisfies the condition~\eqref{eq:lyapunov_cond_exp}, it needs no modification so that $f^*(x,u) = f(x,u)$ for all $u\in\mathcal{U}$; on the other hand, if it violates the condition, we minimally adjust it (in the sense of $\ell^2$ norm) so that the projected model $f^*$ barely satisfies the condition~\eqref{eq:lyapunov_cond_exp} with the equality. This projected model is then guaranteed to be globally exponentially stable at the origin with the feedback controller $u^*$ by construction, as described in Section~\ref{sec:stability}.

We employ neural networks to represent the nominal dynamics model $\hat{f}$, the feedback controller $u^*$, and the Lyapunov function $V$. To ensure that the conditions required for each of these components are met, we carefully choose the architectures of their corresponding neural networks, as described next.

\subsection{Nominal Dynamics Model and Controller}
First, we should make sure that the origin is an equilibrium point for the closed-loop system $f^*_{u^*}$, i.e., $f^*(0, u^*(0)) = \hat{f}(0, u^*(0)) = 0$. 
This could be ensured by setting the nominal dynamics model as
\begin{equation}\label{eq:fhat}
    \hat{f}(x,u) = g_f(x,u) - g_f(0, u^*(0))
\end{equation}
where $g_f:\mathbb{R}^n\times\mathbb{R}^m\rightarrow\mathbb{R}^n$ is an unconstrained neural network with an arbitrary architecture.

For the controller, we require a candidate controller to have its range confined to a pre-specified set $\mathcal{U}$. The control inputs are often generated by motors in most robotics applications and thus are saturated by physical limitations. To handle such prevalent scenarios, we consider $\mathcal{U}=\{u\in\mathbb{R}^m: -u_{\lim}\leq u\leq u_{\lim}\}$ for some $u_{\lim}\geq 0$. Then, we limit the range of the feedback controller by setting 
\begin{equation}\label{eq:ustar}
    u^*(x) = \mathsf{diag}(u_{\lim})\tanh(g_u(x)),
\end{equation}
where $\mathsf{diag}(u_{\lim})$ is a diagonal matrix with its diagonal being $u_{\lim}$, and $g_u:\mathbb{R}^n\rightarrow\mathbb{R}^m$ is an unconstrained neural network.

\subsection{Lyapunov Function}
The eligibility conditions for the Lyapunov function are threefold. First, $V$ should be positive definite, i.e., $V(x) > 0$ for $x \neq 0$ and $V(0) = 0$. We ensure this condition by defining $V$ in the form of
\begin{equation}\label{eq:psd}
    V(x) = \sigma(g_V(x) - g_V(0)) + \epsilon_{\mathrm{pd}} \|x\|^2
\end{equation}
for some $\epsilon_{\mathrm{pd}}>0$, where $g_V:\mathbb{R}^n\rightarrow\mathbb{R}$ is a neural network and $\sigma:\mathbb{R}\rightarrow\mathbb{R}$ is a function with $\sigma(w)=0$ for all $w\leq 0$.

Further, $V$ should be continuously differentiable. We secure this property by using the smoothed ReLU function
\begin{equation}
    \sigma(z) = \begin{cases}
        0 & \text{if } x\leq 0\\
        x^2/2d & \text{if } 0<x<d\\
        x-d/2 & \text{otherwise}
    \end{cases}
\end{equation}
in~\eqref{eq:psd} and as the activations for the neural network $g_V$.

Lastly, $V$ should not have non-zero critical points since its time derivative is strictly negative for any non-zero state. One way to ensure this property is using an input-convex neural network (ICNN) for $g_V$ as introduced by \citet{ManekKolter2019}. Together with the strictly increasing $\sigma$, $V$ is strictly convex and free of non-zero critical points. However, Lyapunov functions are not necessarily convex, and using an ICNN restricts the model capacity of $V$. Instead, we consider using an unconstrained neural network for $g_V$ and approximate the projection in~\eqref{eq:projection} as
\begin{equation}\label{eq:projection_approx}
    \hat{f}(x,u) - \nabla V(x)\frac{\mathsf{ReLU}\bigl(\nabla_{\hat{f}}V(x, u^*(x)) + \alpha V(x)\bigr)}{\max(\|\nabla V(x)\|_2^2, \epsilon_{\mathrm{proj}})}
\end{equation}
for some small $\epsilon_{\mathrm{proj}}>0$. Once the network has been trained, \eqref{eq:lyapunov_cond_exp} with positive definite $V(x)$ would imply $\nabla V(x)$ to be non-zero due to the strictly negative $\nabla_{f^*_{u^*}}V(x)$. Then, the approximation~\eqref{eq:projection_approx} would also recover the original projection.

\subsection{Loss Function}
We jointly learn the three components by minimizing the following loss function:
\begin{equation} \label{eq:loss}
    \mathcal{L}(\theta) = \dfrac{1}{N}\sum_{(x,u,\dot{x})\in\mathcal{D}} k(u, u^*(x)) \|\dot{x} - f^*(x,u)\|^2
    + \lambda \|\theta\|_2^2
\end{equation}
where $\theta$, $k(\cdot,\cdot)$, and $\lambda$ denote the parameters for the neural networks, a kernel function, and the regularization coefficient, respectively. The kernel function $k:\mathcal{U}\times\mathcal{U}\rightarrow\mathbb{R}$ attains a larger value as the two arguments are getting closer. By weighting the $\ell^2$ difference, we encourage $u^*$ to learn a policy that generates trajectories with smaller model error. Conversely, we also encourage $f^*$ to learn better around feasible trajectories, i.e., $(x, u^*(x))$.
In this work, we use the following kernel function:
\begin{equation}
    k(u, u') = 1 + \exp(-\beta\|u-u'\|_2^2).
\end{equation}


\section{Stability Guarantees}\label{sec:stability}
In this section, we rigorously analyze the stability properties of the learned models. We first establish the inherent stability of the constructed model $f^*$ with the feedback controller $u^*$ (Theorem~\ref{thm:fstar}). Then, we connect the inherent stability to the behavior of the true system $f$ depending on the learning performance (Theorem~\ref{thm:f}); see Appendix~\ref{app:proof_fstar}-\ref{app:proof_f} for proofs.

\begin{theorem}[Inherent Stability]\label{thm:fstar}
    Consider a closed-loop system $f^*_{u^*}$ for a projected dynamics $f^*$ defined by \eqref{eq:fstar} with $\hat{f}$ in~\eqref{eq:fhat}, $u^*$ in~\eqref{eq:ustar}, and $V$ in~\eqref{eq:psd}. For any $r_2\geq r_1>0$, let $B_{r_1,r_2} := \{x\in\mathcal{X}:r_1\leq\|x\|_2\leq r_2\}$. Then,\\
    1) $f^*_{u^*}$ is exponentially stable at the origin w.r.t.  $B_{r_1,r_2}$.\\
    2) $f^*_{u^*}$ is exponentially stable at the origin w.r.t. $\mathcal{X}$ if there exists $c>0$ such that $V(x)\leq c\|x\|_2^2\,$ for all $x\in\mathcal{X}$.
\end{theorem}

\coils{} inherently guarantees the exponential stability of the learned models, as proven in Theorem~\ref{thm:fstar}. Note that this property holds even for any initialization of the models. Next, we further argue about its connection to the true dynamics. We could expect this stability property to be transferred to the true dynamics when satisfactory learning performance is achieved. We elaborate on this relationship in the next theorem.

\begin{theorem}\label{thm:f}
    Consider a dynamical system $f$ in~\eqref{eq:system} and its learned model $f^*$ in~\eqref{eq:fstar} with $\hat{f}$ in~\eqref{eq:fhat}, $u^*$ in~\eqref{eq:ustar}, and $V$ in~\eqref{eq:psd}. Assume $f$ and $f^*$ are Lipschitz continuous with Lipschitz constants $L_f$ and $L_{f^*}$, respectively. Let $\delta$ and $e$ denote how densely and accurately, respectively, $f^*$ is learned by defining
    \begin{equation}
    \begin{aligned}
        \delta &:= \sup_{x\in\mathcal{X}} \min_{(y,v)\in\mathcal{D}} \|(x,u^*(x))-(y,v)\|_2,\\
        e &:= \max_{(y,v)\in\mathcal{D}} \|f(y,v) - f^*(y,v)\|_2.
    \end{aligned}
    \end{equation}
    For any $r>0$, let the neighborhood of the origin $N_r:=\{x\in\mathcal{X}: \|x\|_2 < r\}$ and 
    $M_r := \sup_{x\in\mathcal{X}\setminus N_r} \|\nabla V(x)\|_2$. Then, the closed-loop system $f_{u^*}$ always arrives at $N_r$, i.e., for any trajectory $x(t)$ generated by $f_{u^*}$ from $x(0)\in\mathcal{X}$, there exists $T\geq 0$ such that $x(T)\in N_r$, if $\delta$ and $e$ are small enough such that
    \begin{equation}\label{eq:learning_cond}
        (L_f + L_{f^*})\delta + e < \dfrac{\alpha\epsilon_{\mathrm{pd}} r^2}{M_r}.
    \end{equation}
\end{theorem}

\section{Experiments}\label{sec:exp}
\begin{table}[t]
    \centering
        \caption{Hyperparameters for the experiments}
    \begin{tabular}{|c|c|}
        \hline
        parameter & value\\
        \hline
        $\alpha$  & 1.0 or 5.0\\
        $\beta$  & $5/u_{\lim}$\\
        $\lambda$ & 0.0\\
        $\epsilon_{\mathrm{pd}}$ & 0.5\\
        $\epsilon_{\mathrm{proj}}$ & 0.001\\
        $d$ & 0.005\\
        \hline
    \end{tabular}
    \label{tab:hyper}
\end{table}
In this section, we demonstrate the effectiveness of \coils{} in stabilizing unknown nonlinear control systems. We first verify the stability properties of the projected models achieved by our proposed architecture (Section~\ref{sec:exp_rand}). Then, we demonstrate the performance and accuracy of the learned controller and dynamics, respectively, in three different control problems (Section~\ref{sec:exp_vanderpol}-\ref{sec:exp_bicycle}).

For each scenario, we train the models with $N=10^5$ data tuples which are uniformly sampled over $\mathcal{X}=\{x\in\mathbb{R}^n:x_{lb}\leq x\leq x_{ub}\}$ and $\mathcal{U}=\{u\in\mathbb{R}^m: -u_{\lim}\leq u\leq u_{\lim}\}$. The models are constructed with neural networks as described in Section~\ref{sec:learning}. We use 3-layer fully connected neural networks (FCN) for $g_f, g_u$, and $g_V$. They have 100, 50, and 50 hidden neurons, respectively, in each hidden layer. For $g_V$, we add $\mathsf{tanh}$ activation multiplied by 10 at the output layer to limit the scale of the Lyapunov function. The models are trained using a mini-batch gradient descent optimizer with gradient clipping and a learning rate 0.0001. The other hyperparameters used in the experiments are presented in Table~\ref{tab:hyper}.

\begin{figure}[t]
    \centering
    \begin{subfigure}
        \centering
        \includegraphics[height=.17\textheight]{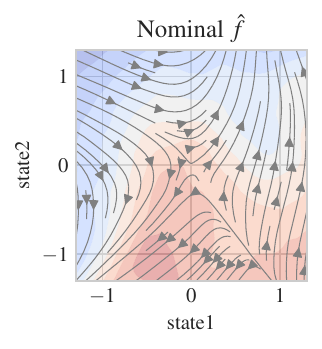}
    \end{subfigure}
    \begin{subfigure}
        \centering
        \includegraphics[height=.17\textheight]{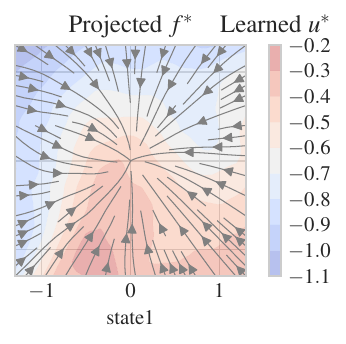}
    \end{subfigure}
    \begin{subfigure}
        \centering
        \includegraphics[height=.14\textheight]{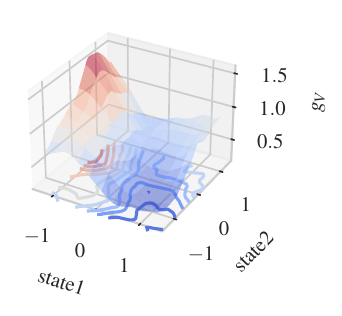}
    \end{subfigure}
    \begin{subfigure}
        \centering
        \includegraphics[height=.14\textheight]{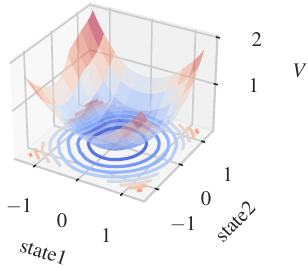}
    \end{subfigure}
    \caption{Closed-loop dynamics and Lyapunov function for randomly initialized models. Top-left: Nominal model $\hat{f}$ with controller $u^*$. Top-right: Projected model $f^*$ with controller $u^*$. Bottom-left: Neural network $g_V$. Bottom-right: Candidate Lyapunov function $V$.}
    \label{fig:rand}
\end{figure}

\subsection{Random Networks}\label{sec:exp_rand}
In this experiment, we investigate the efficacy of the projection with randomly initialized models. As proven in Theorem~\ref{thm:fstar}, we verify that the projected model $f^*$ in~\eqref{eq:fstar} equips the inherent stability in closed-loop with the feedback controller $u^*$ even for their random initialization. We initialize the neural networks for $\mathcal{X}\in\mathbb{R}^2$ and $\mathcal{U}\in\mathbb{R}$. The results are shown in Figure~\ref{fig:rand}. The projected model $f^*$ is stabilized at the origin by the controller $u^*$, while the nominal model $\hat{f}$ is not. As induced by~\eqref{eq:fhat}, the nominal model $\hat{f}$ as well as the projected model $f^*$ achieves zero value at the origin in closed-loop with $u^*$. This ensures that the origin is an equilibrium point of the closed-loop systems. Interestingly, we observe that the candidate Lyapunov function $V$ does not have any critical points although we use a generic FCN for $g_V$ in~\eqref{eq:psd}, as shown in the figure. This favored construction is observed in most cases. 

\begin{figure}[t]
    \centering
    \begin{subfigure}
        \centering
        \includegraphics[height=.17\textheight]{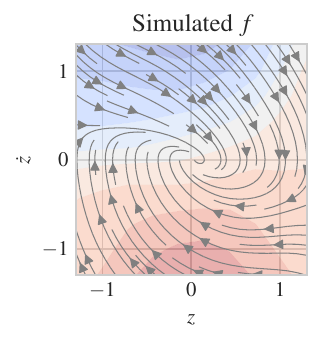}
    \end{subfigure}
    \begin{subfigure}
        \centering
        \includegraphics[height=.17\textheight]{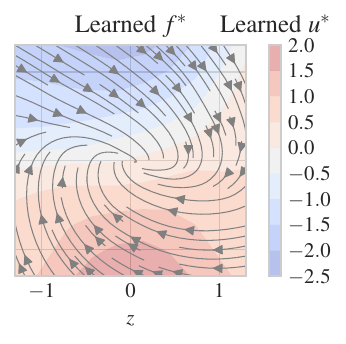}
    \end{subfigure}
    \begin{subfigure}
        \centering
        \includegraphics[height=.15\textheight]{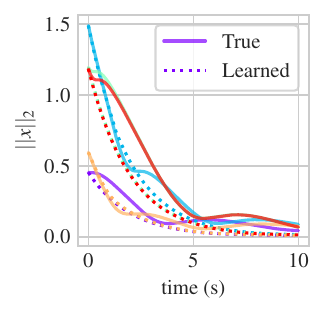}
    \end{subfigure}
    \begin{subfigure}
        \centering
        \includegraphics[height=.15\textheight]{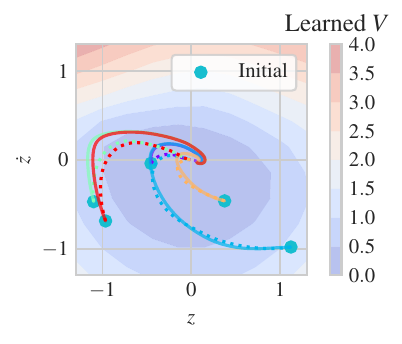}
    \end{subfigure}
    \caption{Comparison of simulation and the learned model for the Van der Pol oscillator. Top-left: Closed-loop dynamics for the true model $f$ with the learned controller $u^*$. Top-right: Closed-loop dynamics for the learned model $f^*$ with the learned controller $u^*$. Bottom-left: Comparison of 5 randomly initialized trajectories for the true and learned system. Bottom-right: Plot of the trajectories in the state space with the contour map of the learned Lyapunov function.}
    \label{fig:cvanderpol_dyn}
\end{figure}

\subsection{Van der Pol Oscillator}\label{sec:exp_vanderpol}
We test our proposed method in three different control problems. We start with stabilizing a system, the Van der Pol oscillator, which is already stable at the origin without any control input. The autonomous Van der Pol oscillator is a well-known nonlinear system that exhibits stable oscillations in the form of a limit cycle. By adding a control input $u\in\mathbb{R}$, we consider the true dynamics as
\begin{equation}
    \Ddot{z} = u - z + \mu (1 - z^2) \dot{z}
\end{equation}
for state $x=[z, \dot{z}]\in\mathbb{R}^2$ with the parameter $\mu{=}1$. The dataset is sampled with $x_{lb} {=} [-1.3,-1.3], x_{ub} {=} [1.3,1.3], u_{\lim} {=} 5$. The results are shown in Figure~\ref{fig:cvanderpol_dyn}. The closed-loop dynamics for the true system $f_{u^*}$ and the learned system $f^*_{u^*}$ show similar behaviors. Due to some model errors, the trajectories are not perfectly aligned, but the learned controller successfully stabilizes both true and learned systems to the origin.

\begin{figure}[t]
    \centering
    \begin{subfigure}
        \centering
        \includegraphics[width=.13\textwidth]{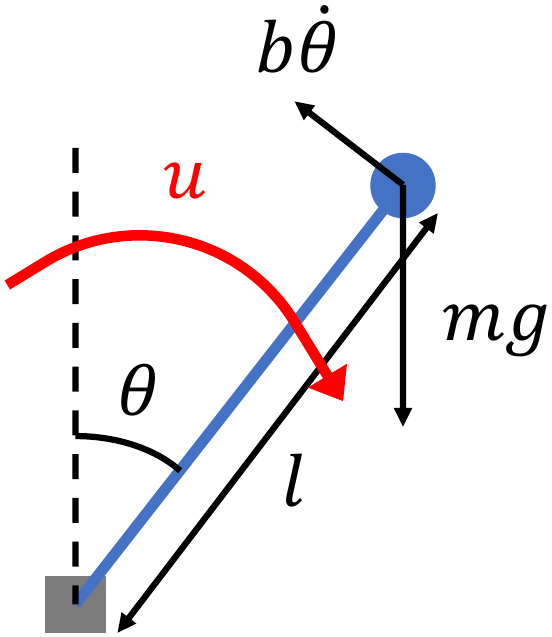}
    \end{subfigure}
    \begin{subfigure}
        \centering
        \includegraphics[width=.19\textwidth]{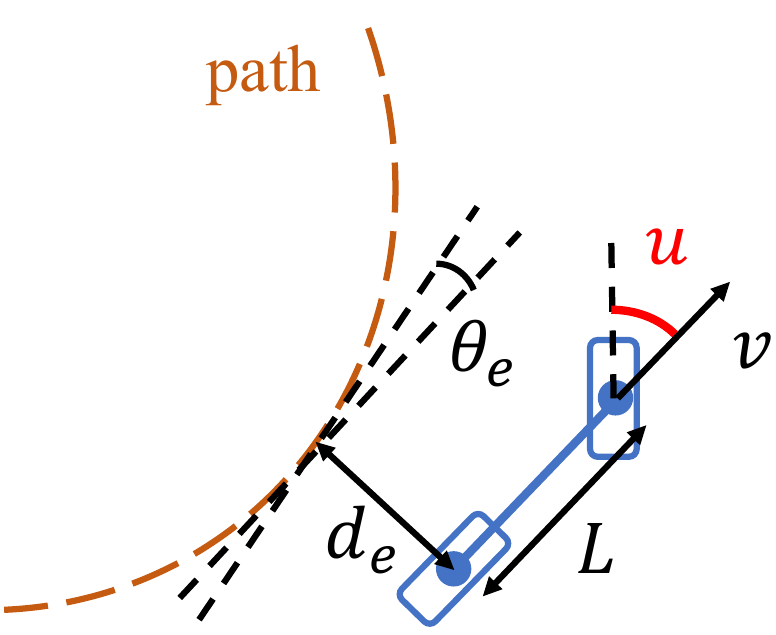}
    \end{subfigure}
    \caption{Schematic diagrams for (left) the inverted pendulum and (right) the bicycle path following system.}
    \label{fig:diagram}
    \vspace{0.6cm}
    \begin{subfigure}
        \centering
        \includegraphics[height=.17\textheight]{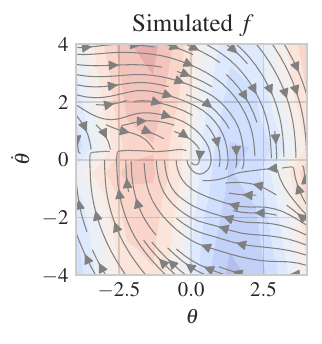}
    \end{subfigure}
    \begin{subfigure}
        \centering
        \includegraphics[height=.17\textheight]{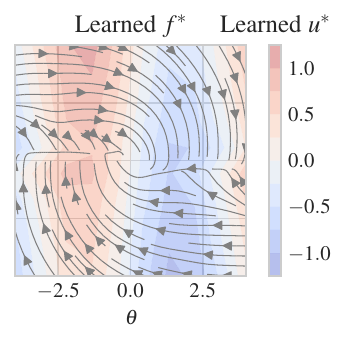}
    \end{subfigure}
    \begin{subfigure}
        \centering
        \includegraphics[height=.15\textheight]{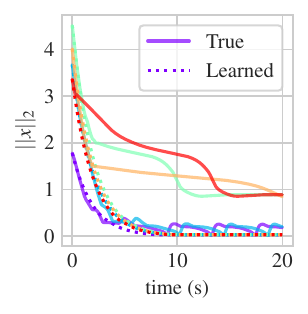}
    \end{subfigure}
    \begin{subfigure}
        \centering
        \includegraphics[height=.15\textheight]{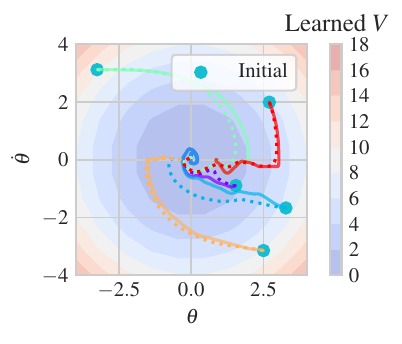}
    \end{subfigure}
    \caption{Comparison of simulation and the learned model for the inverted pendulum system. Top-left: Closed-loop dynamics for the true model $f$. Top-right: Closed-loop dynamics for the learned model $f^*$. Bottom-left: Comparison of 5 randomly initialized trajectories. Bottom-right: Plot of the trajectories in the state space with the contour map of the learned Lyapunov function.}
    \label{fig:ipendulum_dyn}
\end{figure}

\subsection{Inverted Pendulum}\label{sec:exp_ipendulum}
Next, we demonstrate \coils{} in a more challenging control problem, stabilizing an inverted pendulum. The inverted pendulum, shown in Figure~\ref{fig:diagram}, easily falls off the origin without proper control due to gravity. For angular position $\theta$ from the inverted position and its angular velocity $\dot{\theta}$, we consider states $x=[\theta, \dot{\theta}]\in\mathbb{R}^2$. By providing a torque $u\in\mathbb{R}$ at the pivot as a control input, we consider the true dynamics as 
\begin{equation}
    \Ddot{\theta} = \frac{mgl\sin{\theta} +u -b\dot{\theta}}{ml^2}.
\end{equation}
We set the parameters $m=0.15, g=9.81, l=0.5, b=0.1$ with
$x_{lb}=[-4,-4], x_{ub}=[4,4], u_{\lim}=5$. The results are shown in Figure~\ref{fig:ipendulum_dyn}. The overall behaviors of the closed-loop dynamics are similar for the true system $f_{u^*}$ and the learned system $f^*_{u^*}$. However, the trajectories exhibit quite dissimilar $\ell^2$ norm evolution. This difference occurs from the model errors in the small area around $\dot{\theta}=0$. The small magnitude of $f_{u^*}$ in those areas slows down the movement of the state even though the direction of the movement is similar for both systems. These model errors could be improved if we utilize the physical relationship between the state elements, i.e., the second element is the derivative of the first element.

\begin{figure}[t]
    \centering
    \begin{subfigure}
        \centering
        \includegraphics[height=.17\textheight]{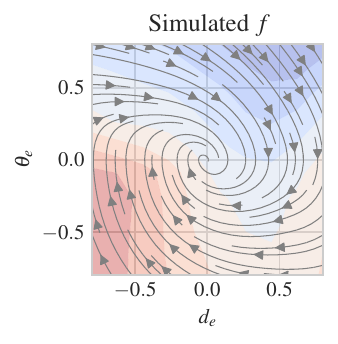}
    \end{subfigure}
    \begin{subfigure}
        \centering
        \includegraphics[height=.17\textheight]{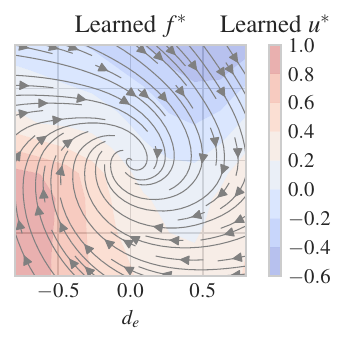}
    \end{subfigure}
    \begin{subfigure}
        \centering
        \includegraphics[height=.15\textheight]{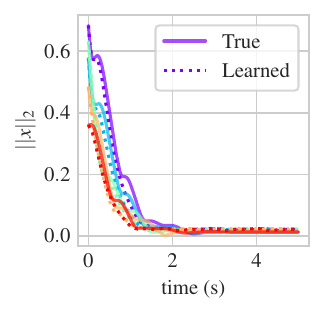}
    \end{subfigure}
    \begin{subfigure}
        \centering
        \includegraphics[height=.15\textheight]{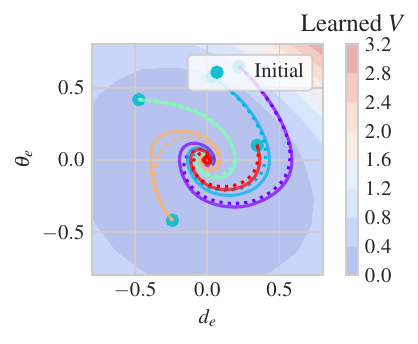}
    \end{subfigure}
    \caption{Comparison of simulation and the learned model for the bicycle path following system. Top-left: Closed-loop dynamics for the true model $f$ with the learned controller $u^*$. Top-right: Closed-loop dynamics for the learned model $f^*$ with the learned controller $u^*$. Bottom-left: Comparison of 5 randomly initialized trajectories for the true and learned system. Bottom-right: Plot of the trajectories in the state space with the contour map of the learned Lyapunov function.}
    \label{fig:bicycle_dyn}
\end{figure}

\subsection{Bicycle Path Following}\label{sec:exp_bicycle}
The previous examples are both control-affine systems, while \coils{} is applicable to general control systems without such structure. In this experiment, we evaluate \coils{} for a control system in which the control input has a nonlinear effect. We consider the problem of controlling a constant-speed bicycle to follow a unit circle path as shown in Figure~\ref{fig:diagram}. We aim to drive the distance error $d_e$ and angular error $\theta_e$ to be zero by controlling the steering angle $u\in\mathbb{R}$. For the state 
$x=[d_e, \theta_e]\in\mathbb{R}^2$, the dynamics are given as
\begin{equation}
\begin{aligned}
    \dot{d_e} &= v\sin{\theta_e},\\
    \dot{\theta_e} &= \frac{v\tan{u}}{L} - \frac{v\cos{\theta_e}}{1-d_e}.
\end{aligned}
\end{equation}
We set the parameters $v=6, L=1$ with $x_{lb}=[-0.8,-0.8]$, $x_{ub}=[0.8,0.8]$, $u_{\lim}=0.4\pi$. The results are shown in Figure~\ref{fig:bicycle_dyn}. The closed-loop dynamics for the true system $f_{u^*}$ and the learned system $f^*_{u^*}$ are similar to each other. Also, the learned controller generates trajectories close for both systems and successfully stabilizes both systems to the origin.

\section{Conclusion}
We presented a new data-driven method \coils{} to stabilize unknown controlled dynamical systems. We jointly learn the dynamics model and a feedback controller with a Lyapunov function, such that the projected model is \emph{guaranteed by construction} to be stabilized in closed-loop by the learned controller. We further showed that, under certain assumptions on the fidelity of our learned dynamics model, the learned controller is also guaranteed to stabilize the true dynamics. We demonstrated the performance of our method in the simulation of a number of controlled nonlinear dynamical systems.

There are several interesting avenues for future work. First, it is feasible to explore different loss functions, for instance, to include control-oriented metrics. This work focused on learning the dynamics model without explicitly optimizing the performance of the controller. Since there exist various controllers that can stabilize the system in practice, guiding the learning to find one with better performance would be an interesting direction. Also, exploring different hyperparameters for our method would be beneficial to further optimize our approach. Furthermore, one can think about extending our approach to partially observable systems. It would be challenging but worthwhile to find an output-feedback controller that stabilizes unknown dynamical systems.





\bibliography{strings,ref}
\bibliographystyle{abbrv}

\appendix

\section{Proofs}

\subsection{Special Case: Control-Affine Systems}
When the true dynamics~\eqref{eq:system} is known to have a control-affine structure, we can learn its Lyapunov control without explicitly learning a feedback controller. Consider a control-affine system in standard form
\begin{equation}
    \dot{x}(t) = f_1(x(t)) + f_2(x(t))u(t)
\end{equation}
for $x(t)\in\mathcal{X}\subset\mathbb{R}^n, u(t)\in\mathcal{U}\subset\mathbb{R}^m$. Exploiting this structural knowledge, we can also form the projected model $f^*$ to be control-affine as well, i.e., $f^*(x,u)=f^*_1(x)+f^*_2(x)u$.
Then, the Lyapunov stability condition~\eqref{eq:lyapunov_cond_exp} becomes
\begin{equation}\label{eq:lyapunov_cond_affine}
    \nabla V(x)^\top (f^*_1(x) + f^*_2(x)u^*(x)) \leq -\alpha V(x)
\end{equation}
for all $x\in\mathcal{X}\setminus\{0\}$ with a given $\alpha>0$. Due to the control-affine structure, we can easily compute the controller that minimizes the LHS of~\eqref{eq:lyapunov_cond_affine} instead of directly learning it:
\begin{equation}\label{eq:ustar_affine}
    u^*(x) = \argmin_{u\in\mathcal{U}} \nabla V(x)^\top f^*_2(x)u.
\end{equation}
Considering $\mathcal{U}=\{u\in\mathbb{R}^m: -u_{\lim}\leq u\leq u_{\lim}\}$ for some $u_{\lim}\geq 0$, \eqref{eq:ustar_affine} is reduced to
\begin{equation}\label{eq:controller_affine}
    u^{*(i)}(x) = - \mathsf{sign}(\{\nabla V(x)^\top f^*_2(x)\}^{(i)}) u_{\lim}^{(i)}
\end{equation}
for each $i=1,2,\dots,m$ where $(\cdot)^{(i)}$ denotes the $i$-th component. Then, we can just optimize the neural networks for the dynamics model and the Lyapunov function as before with the induced feedback controller~\eqref{eq:controller_affine}.

Constructing the projected model $f^*$ to be control-affine can be simply done by setting the nominal model to be control-affine. Specifically, $\hat{f}(x,u) = \hat{f_1}(x) + \hat{f_2}(x)u$ induces $f^*_2 = \hat{f}_2$. To ensure that the origin is an equilibrium for this structure as in~\eqref{eq:fhat}, we can set
\begin{equation}
\begin{aligned}
    \hat{f_2}(x) &= g_{f_2}(x),\\
    \hat{f_1}(x) & = g_{f_1}(x) - g_{f_1}(0) - \hat{f_2}(0) u^*(0),
\end{aligned}
\end{equation}
where $g_{f_1}:\mathbb{R}^n\rightarrow\mathbb{R}^n$ and $g_{f_2}:\mathbb{R}^n\rightarrow\mathbb{R}^{n\times m}$ are unconstrained neural networks.

\subsection{Proof of \cref{thm:fstar}} \label{app:proof_fstar}

\proof
We prove the exponential stability of the closed-loop system $f^*_{u^*}$ by deducing an exponentially decaying upper bound of $\|x(t)\|_2$ from that of the Lyapunov function $V(x(t))$.

First, we compute the upper bound of $V$ from the Lyapunov condition~\eqref{eq:lyapunov_cond_exp} which holds for the projected model.
Let $x(t)$ be a trajectory generated by the closed-loop system $f^*_{u^*}$ with an initial state $x(0)\in\mathcal{X}$. Then, for all $t\geq 0$,
\begin{equation}
\begin{aligned}
    \dfrac{d}{dt} V(x(t)) &= \nabla V(x(t))^\top f^*(x(t), u^*(x(t)))\\
    & \leq -\alpha V(x(t)).
\end{aligned}
\end{equation}
By integrating this inequality, we get the upper bound as
\begin{equation}\label{eq:V_upper}
    V(x(t)) \leq V(x(0))e^{-\alpha t}.
\end{equation}

Next, we deduce an exponentially decaying upper bound of $\|x(t)\|_2$ from~\eqref{eq:V_upper}. From the positive definite structure of $V$ in~\eqref{eq:psd}, $V(x)\geq\epsilon_{\mathrm{pd}} \|x\|_2^2$, which implies
\begin{equation}\label{eq:x_upper}
    \|x(t)\|_2
    \leq \sqrt{\dfrac{V(x)}{\epsilon_{\mathrm{pd}}}}
    \leq\sqrt{\dfrac{V(x(0))e^{-\alpha t}}{\epsilon_{\mathrm{pd}}}}.
\end{equation}
Then, the remaining step is expressing the upper bound in terms of $\|x(0)\|_2$. Given $r_2\geq r_1>0$, we define
\begin{equation}\label{eq:sup}
    M = \sup_{x\in\mathbb{R}^n} \dfrac{V(x)}{\|x\|_2^2} \,\,\text{ s.t. } r_1\leq\|x\|_2\leq r_2.
\end{equation}
The feasible set of the optimization in~\eqref{eq:sup} is closed and bounded, hence compact. Then, the supremum $M$ is attained on the feasible set and, thus, is finite. Since the feasible set includes $B_{r_1,r_2}$, \eqref{eq:sup} implies that for any $x(0)\in B_{r_1,r_2}$, $V(x(0))\leq M\|x(0)\|_2^2$. Together with~\eqref{eq:x_upper},
\begin{equation}
    \|x(t)\|_2\leq\sqrt{\dfrac{V(x(0))e^{-\alpha t}}{\epsilon_{\mathrm{pd}}}}
    \leq \sqrt{\dfrac{M}{\epsilon_{\mathrm{pd}}}}\|x(0)\|_2 e^{-\alpha t/2}.
\end{equation}
Thus, the system $f^*_{u^*}$ is exponentially stable at the origin w.r.t. $B_{r_1,r_2}$.

If there exists $c>0$ such that $V(x)\leq c\|x\|_2^2$ for all $x\in\mathcal{X}$, then similarly (or by \cref{prop:exp}), the system $f^*_{u^*}$ is exponentially stable at the origin w.r.t. $\mathcal{X}$.
\qed

\subsection{Proof of \cref{thm:f}} \label{app:proof_f}

\proof
Suppose \eqref{eq:learning_cond} holds and represent the difference as
\begin{equation} \label{eq:diff}
    \Delta := \dfrac{\alpha\epsilon_{\mathrm{pd}} r^2}{M_r} - (L_f + L_{f^*})\delta - e >0.
\end{equation}
We first prove that $V$ is strictly decreasing along any trajectories outside $N_r$ for the true closed-loop system $f_{u^*}$. Then, we show that they arrive at $N_r$ by contradiction.

For all $x\in\mathcal{X}\setminus N_r$, $V$ is strictly decreasing for the closed-loop system $f^*_{u^*}$ since
\begin{equation}
    \nabla_{f^*_{u^*}} V(x) \leq - \alpha V(x) \leq - \alpha \epsilon_{\mathrm{pd}} \|x\|^2 \leq - \alpha \epsilon_{\mathrm{pd}} r^2 < 0.
\end{equation}
Then, we show $V$ is also strictly decreasing for the true system $f_{u^*}$ by bounding the difference
\begin{equation}
\begin{aligned}
    &\nabla_{f_{u^*}} V(x) - \nabla_{f^*_{u^*}} V(x)\\
    &\hspace{0.8cm}\leq \|\nabla V(x)\|_2 \|f(x, u^*(x)) - f^*(x, u^*(x))\|_2.
\end{aligned}
\end{equation}
The generalization error of the learned closed-loop system can be bounded as
\begin{equation}
\begin{aligned}
    &\|f(x,u^*(x))-f^*(x,u^*(x))\|_2 \\
    &\leq \|f(x,u^*(x)) - f(y,v)\|_2 + \|f(y,v) - f^*(y,v)\|_2\\
    &\hspace{1.5cm} + \|f^*(y,v) - f^*(x,u^*(x))\|_2\\
    &\leq L_f\delta + e + L_{f^*}\delta = (L_f + L_{f^*})\delta + e
\end{aligned}
\end{equation}
where $(y,v)$ is the nearest data sample in $\mathcal{D}$ from $(x,u^*(x))$. The first inequality is from the triangle inequality and the second inequality is from the definitions of $L_f, L_{f^*}, \delta, e$. Putting the above inequalities together with \eqref{eq:diff}, we have
\begin{equation}\label{eq:lyap_f}
\begin{aligned}
    \nabla_{f_{u^*}} V(x) &\leq \nabla_{f^*_{u^*}} V(x) \\
    & \hspace{0.5cm}+\! \|\nabla V(x)\|_2 \|f(x, u^*(x)) \!-\! f^*(x, u^*(x))\|_2\\
    &\leq - \alpha \epsilon_{\mathrm{pd}} r^2 + M_r((L_f + L_{f^*})\delta + e)
    = -M_r \Delta.
\end{aligned}
\end{equation}

Now, we prove that all trajectories arrive at $N_r$ by contradiction. Suppose there exists $x(0)\in\mathcal{X}\setminus N_r$ such that the closed-loop system $f_{u^*}$ generates a trajectory $x(t)\notin N_r$ for all $t\geq 0$. Then, $\|x(t)\|_2\geq r$ implies 
\begin{equation}\label{eq:contradict}
    V(x(t))\geq \alpha\epsilon_{\mathrm{pd}} r^2.
\end{equation}
However, since $x(t)\notin N_r$, $\dot{V}(x(t))\leq-M_r \Delta<0$ for all $t\geq 0$. Through the integration from $V(x(0))$,
\begin{equation}
\begin{aligned}
    V(x(t)) &= V(x(0)) + \int_0^t \dot{V}(x(\tau)) d\tau\\
    &\leq V(x(0)) -M_r \Delta t.
\end{aligned}
\end{equation}
This implies that there exists large $T$ such that $V(x(T)) < \alpha\epsilon_{\mathrm{pd}} r^2$ which is a contradiction to~\eqref{eq:contradict}. Thus, $x(t)$ arrives at $N_r$.
\qed








\end{document}